\documentclass[12pt]{iopart}
\usepackage{iopams}  
\usepackage{graphicx}

\begin{document}

\eqnobysec

\newtheorem{proposition}{Proposition}[section]
\renewcommand\vec[1]{\boldsymbol{#1}}
\renewcommand\Re{\mathop{\mbox{\rm Re}}}
\renewcommand\Im{\mathop{\mbox{\rm Im}}}

\newcommand\Shift[1]{\mathbb{T}_{#1}}
\newcommand\Shifted[3][0]{\ifcase #1
   \Shift{#2}#3        \or   %0: direct
   (\Shift{#2}#3)      \or   %1: direct+braces
   \Shift{#2}^{-1}#3   \or   %2: inverse
   (\Shift{#2}^{-1}#3) \else %3: inverse+braces
   #3[#2] 
   \fi} 
\newcommand\ShiftR[1]{\mathbb{T}^{\scriptscriptstyle{R}}_{#1}}

\newcommand\A[1]{\ifcase#1 A \or A_{1} \or A_{2} \else ? \fi} 
\renewcommand\L[1]{\ifcase#1 L \or L_{1} \or L_{2} \else ? \fi} 
\newcommand\G[1]{\ifcase#1 G \or G_{1} \or G_{2} \else ? \fi} 
\newcommand\keta[1]{ \if ?#1? |1\rangle \else |\ell_{#1}\rangle\fi} 
\newcommand\braa[1]{\langle a_{#1}|} 
\newcommand\brab[1]{\langle\beta_{#1}|} 
\renewcommand\u[1]{K_{#1}} 
\newcommand\q[1]{\ifcase#1 p \or q \or r \or s \or w \else ? \fi}

%%%%%%%%%%%%%%%%%%%%%%%%%%%%%%%%%%%%%%%%%%%%%%%%%%%%%%%%%%%%%%%%%%%%%%%%%%%%%%%%
%%%%%%%%%%%%% Title, author, address etc. %%%%%%%%%%%%%%%%%%%%%%%%%%%%%%%%%%%%%%
%%%%%%%%%%%%%%%%%%%%%%%%%%%%%%%%%%%%%%%%%%%%%%%%%%%%%%%%%%%%%%%%%%%%%%%%%%%%%%%%

\title{Solitons of the vector KdV and Yamilov lattices.}
\author{V.E. Vekslerchik}
\address{
Usikov Institute for Radiophysics and Electronics \\
12, Proskura st., Kharkov, 61085, Ukraine 
}
\ead{vekslerchik@yahoo.com}

\begin{abstract}
We study a vector generalizations of the lattice KdV equation and one of 
the simplest Yamilov equations. 
We use algebraic properties of a certain class of matrices 
to derive the $N$-soliton solutions. 
\end{abstract}

\submitto{\JPA}
%

%%%%%%%%%%%%%%%%%%%%%%%%%%%%%%%%%%%%%%%%%%%%%%%%%%%%%%%%%%%%%%%%%%%%%%%%%%%%%%%
\section{Introduction.}
%%%%%%%%%%%%%%%%%%%%%%%%%%%%%%%%%%%%%%%%%%%%%%%%%%%%%%%%%%%%%%%%%%%%%%%%%%%%%%%

We study a generalization of two well-known equations, 
the lattice KdV equation, 
\begin{equation}
  \left( u_{m+1,n+1} - u_{m,n} \right)
  \left( u_{m+1,n} - u_{m,n+1} \right)
  = 
  1, 
\label{eq:kdv}
\end{equation}
and one of the equations from the Yamilov list,
\begin{equation}
  \frac{d u_{n}}{d t} 
  = 
  \frac{ 1 }
       { u_{n+1} - u_{n-1} }.
\label{eq:nve}
\end{equation}

Equation \eref{eq:kdv} has been introduced by Capel, Nijhoff and coauthors 
\cite{CWN86,PNC90,CNP91,NC95} and now is often referred to as equation H1 
from the Adler--Bobenko--Suris list \cite{ABS03}. During its more than 30-year 
history it has attracted much attention and is one of the most-studied discrete 
integrable systems. 

Equation \eref{eq:nve}, sometimes referred to as Yamilov discretization of the 
Krichever--Novikov equation, is known since the work by Yamilov \cite{Y83} who 
classified all integrable semi-discrete equations of the form 
$du_{n}/dt = f(u_{n-1}, u_{n}, u_{n+1})$ using the generalized symmetry method 
(see also \cite{Y06,MSY87}). 
Equation \eref{eq:nve} is related to the the well-known Volterra equation. 
It has been shown in \cite{PV03} that it describes the simplest negative flow 
of the Volterra hierarchy.

Despite their different appearance, equations \eref{eq:kdv} and \eref{eq:nve} 
are known to be closely related. For example, it has been demonstrated in 
\cite{LPSY08} that generalized symmetries of \eref{eq:kdv} are described by 
\eref{eq:nve}. In other words, equation \eref{eq:kdv} can be viewed as 
describing the B\"acklund transformations of equation \eref{eq:nve}.

The models we discuss here are 
\begin{equation}
  \left\| \vec{\phi}_{m+1,n+1} - \vec{\phi}_{m,n} \right\| 
  \left\| \vec{\phi}_{m+1,n} - \vec{\phi}_{m,n+1} \right\| 
  = 
  1
\label{eq:vkdv}
\end{equation}
and 
\begin{equation}
  \frac{d}{d t} \, \vec{\phi}_{n} 
  = 
  \frac{ \vec{\phi}_{n+1} - \vec{\phi}_{n-1} }
       { \left\|\vec{\phi}_{n+1} - \vec{\phi}_{n-1}\right\|^{2} }. 
\label{eq:vnve}
\end{equation}

Here and in what follows the vectors 
$\vec\phi$ are 3-dimensional real vectors,  
$\vec\phi = \left( \phi_{1}, \phi_{2}, \phi_{3} \right)^{T} \in \mathbb{R}^{3} $, 
and $\left\| \vec{\phi} \right\|$ denotes the standard Euclidean norm in 
$\mathbb{R}^{3}$, 
$\left\| \vec{\phi} \right\|^{2} = \sum_{i=1}^{3} \phi_{i}^{2}$. 

Equations \eref{eq:vkdv} and \eref{eq:vnve} can be viewed as `vectorizations' 
of \eref{eq:kdv} and \eref{eq:nve} alternative to ones discussed in 
\cite{CWN86} (compare equations \eref{eq:vkdv} and \eref{eq:vnve} with equations 
(6.9) and (8.5) from \cite{CWN86}).

In this paper, we do not discuss the questions related to the integrability 
of equations \eref{eq:vkdv} and \eref{eq:vnve} such as Lax representation, 
conservation laws, Hamiltonian structures \textit{etc}. 
We restrict ourselves with the problem of finding some \emph{particular} 
solutions, namely the $N$-soliton ones. 

In the next section we introduce an auxiliary system 
which is closely related to the equations we want to solve.  
In section \ref{sec:matrices} we derive some solutions for this system 
using the straightforward calculations involving the soliton matrices 
discussed in \cite{V15}. These solutions are used in section \ref{sec:solitons} 
to construct the $N$-soliton solutions for equations \eref{eq:vkdv} and 
\eref{eq:vnve}.

%%%%%%%%%%%%%%%%%%%%%%%%%%%%%%%%%%%%%%%%%%%%%%%%%%%%%%%%%%%%%%%%%%%%%%%%%%%%%%%
\section{Auxiliary system. \label{sec:aux} }
%%%%%%%%%%%%%%%%%%%%%%%%%%%%%%%%%%%%%%%%%%%%%%%%%%%%%%%%%%%%%%%%%%%%%%%%%%%%%%%

To derive the soliton solutions we start from the bilinear \emph{difference} 
vector equation 
\begin{equation}
  \Shifted{\xi}{ \vec{\phi}} - \Shifted{\eta}{ \vec{\phi}} 
  = 
  \varepsilon_{\xi\eta} 
  \frac{ \Shifted{\xi\eta}{\vec{\phi}} - \vec{\phi} }
       { \left\| \Shifted{\xi\eta}{\vec{\phi}} - \vec{\phi} \right\|^{2} }
\label{eq:dvnve} 
\end{equation}
where $\varepsilon_{\xi\eta}$ is some skew-symmetric constant, 
$\varepsilon_{\xi\eta}=-\varepsilon_{\eta\xi}$ which we introduce to ensure 
the proper symmetry with respect to the interchange of $\xi$ and $\eta$.
\marginpar{OK?} %%%%%%%%%%%%%%%%%%%%%%%%%%%%%%%%%%%%%%%
The symbols $\Shift{\xi}$ stand for the shifts, 
which can be viewed as a generalization of the translations 
$\vec{\phi}(x) \to \vec{\phi}(x+\delta(\xi))$ with some analytic function 
$\delta(\xi)$ and whose particular implementation in our case is specified 
below (see \eref{evol:A}) while the double indices denote combined action of 
different shifts, $\Shift{\xi\eta} = \Shift{\xi}\Shift{\eta}$.

It is easy to show that each solution for \eref{eq:dvnve} provide a solution 
for both \eref{eq:vkdv} and \eref{eq:vnve}.
Indeed, taking the norm of both sides of \eref{eq:dvnve} one immediately 
arrives at 
\begin{equation}
  \left\| \Shifted{\xi\eta}{\vec{\phi}} - \vec{\phi} \right\| \;
  \left\| \Shifted{\xi}{\vec{\phi}} - \Shifted{\eta}{\vec{\phi}} \right\| 
  = 
  |\varepsilon_{\xi\eta}|. 
\label{eq:nphi} 
\end{equation}
Thus, any solution for \eref{eq:dvnve} solves at the same time the equation 
which is (up to a constant in the right-hand side) 
nothing but the difference version of \eref{eq:vkdv}. 
This means that solutions for \eref{eq:dvnve} can be converted, 
by fixing the values $\xi$ and $\eta$, into ones for \eref{eq:vkdv}.

On the other hand, it is easy to check that after applying $\Shift{\eta}^{-1}$ 
and taking the $\xi \to \eta$ limit one arrives at 
\begin{equation}
  \mathbb{D}_{\eta} \vec{\phi} 
  = 
  \frac{ \Shifted{\eta}{\vec{\phi}} - \Shifted[2]{\eta}{\vec{\phi}} }
       { \left\| 
           \Shifted{\eta}{\vec{\phi}} - \Shifted[2]{\eta}{\vec{\phi}} 
         \right\|^{2} } 
\label{eq:dphi} 
\end{equation}
where $\mathbb{D}_{\eta}$ is the differential operator defined as 
\begin{equation}
  \mathbb{D}_{\eta} 
  = 
  \lim_{\xi\to\eta} 
  \frac{ 1 }{ \varepsilon_{\xi\eta} } 
  \left( \Shift{\xi}\Shifted{\eta}^{-1} - 1 \right) 
\label{def:D} 
\end{equation}
(note that the fact that 
$\varepsilon_{\xi\eta}=-\varepsilon_{\eta\xi}$ 
together wih the assumption of analytical dependence of 
$\varepsilon_{\xi\eta}$ on $\xi$ and $\eta$ yields 
$\varepsilon_{\eta\eta}=0$).

Of course, the correspondence between solutions of 
\eref{eq:vkdv}, \eref{eq:vnve} 
(or even their difference versions \eref{eq:nphi} and \eref{eq:dphi}) 
and \eref{eq:dvnve} is not one-to-one. 
Each solution for \eref{eq:dvnve} satisfies \eref{eq:nphi} but the reverse 
statement is not true. The similar situation is with \eref{eq:dvnve} and 
\eref{eq:dphi}. 
However, the fact that using \eref{eq:dvnve} we actually make a reduction 
is not crucial for our consideration because the aim of this work is to derive 
the soliton solutions, a set of \emph{particular} solutions, and, as is shown in 
what follows, the soliton solutions stand this reduction. 

Comparison of the equations \eref{eq:nphi} and \eref{eq:dphi} with 
\eref{eq:kdv} and \eref{eq:nve} suggests the 
following way to derive solutions for the last two equations using the ones for 
\eref{eq:nphi} and \eref{eq:dphi}: 
to identify the shits corresponding to some fixed parameter, say, 
$\mu$ and $\nu$ with the 
translations $m \to m+1$ and $n \to n+1$, 
and to introduce the $t$-dependence in such a way that the action of 
$\mathbb{D}_{\nu}$ defined in terms of the $\Shift{}$-shifts leads to the same 
results as the differentiating with respect to $t$. 
Thus, we set 
\begin{equation}
  \Shifted{\mu}{\vec{\phi}_{m,n}} 
  = 
  \vec{\phi}_{m+1,n},
  \qquad
  \Shifted{\nu}{\vec{\phi}_{m,n}} 
  = 
  \vec{\phi}_{m,n+1}
\end{equation}
for equation \eref{eq:kdv} and 
\begin{equation}
  \Shifted{\nu}{\vec{\phi}_{n}} 
  = 
  \vec{\phi}_{n+1},
  \qquad
  \mathbb{D}_{\nu} \vec{\phi}_{n} 
  = 
  \frac{\partial}{\partial t} \vec{\phi}_{n} 
\end{equation}
for equation \eref{eq:nve}.

Rewriting \eref{eq:dvnve} as a system 
\begin{equation}
  \begin{array}{l}
  A_{\xi\eta} 
  \left( \Shifted{\xi\eta}{\vec{\phi}} - \vec{\phi} \right) 
  = 
  f_{\xi\eta} 
  \left( \Shifted{\xi}{\vec{\phi}} - \Shifted{\eta}{\vec{\phi}} \right) 
  \\[2mm] 
  B_{\xi\eta} 
  \left\| \Shifted{\xi\eta}{\vec{\phi}} - \vec{\phi} \right\|^{2} 
  = 
  f_{\xi\eta} 
  \end{array}
\label{syst:tosolve}
\end{equation}
where new constants $A_{\xi\eta}$ and $B_{\xi\eta}$ satisfy  
\begin{equation}
  A_{\xi\eta} = - A_{\eta\xi}, \quad
  B_{\xi\eta} =  B_{\eta\xi}, \quad
  \frac{ A_{\xi\eta} }{ B_{\eta\xi} } 
  = \varepsilon_{\xi\eta} 
\end{equation}
one can note that the first equation of this system is nothing but 
the difference vector Moutard equation which can be tackled in a standard way. 
Indeed,  the substitutions 
\begin{equation}
  \vec{\phi} = \frac{1}{\tau} \vec{\omega}, 
  \qquad
  f_{\xi\eta} 
  = 
  \frac{ \Shifted[1]{\xi}{\tau} \Shifted[1]{\eta}{\tau} } 
       { \tau \Shifted[1]{\xi\eta}{\tau} }
\end{equation}
lead to the well-known bilinear equation 
\begin{equation}
  A_{\xi\eta} 
  \left(
    \tau 
    \Shifted[1]{\xi\eta}{ \vec{\omega}} 
    - 
    \Shifted[1]{\xi\eta}{ \tau} 
    \vec{\omega} 
  \right) 
  = 
  \Shifted[1]{\eta}{ \tau} 
  \Shifted[1]{\xi}{ \vec{\omega}} 
  - 
  \Shifted[1]{\xi}{ \tau} 
  \Shifted[1]{\eta}{ \vec{\omega}} 
\label{eq:bimoutard}
\end{equation}
which, for example, is the zero-curvature representation of the Miwa equation 
\cite{M82} and whose soliton solutions can be derived, say, by means of the 
Hirota approach. 

However, to satisfy the second equation from \eref{syst:tosolve} turns out to 
be a non-trivial problem. The main difficulty arises from the fact that, 
contrary to equation \eref{eq:bimoutard}, it is not a bilinear one. 
In terms of $\vec{\omega}$, we arrive at a \emph{quadrilinear} equation 
\begin{equation}
  B_{\xi\eta} 
  \left\| 
    \tau \Shifted{\xi\eta}{\vec{\omega}} 
    - 
    \Shifted[1]{\xi\eta}{\tau} \vec{\omega} 
  \right\|^{2}
  = 
  \tau 
  \Shifted[1]{\xi}{\tau} 
  \Shifted[1]{\eta}{\tau} 
  \Shifted[1]{\xi\eta}{\tau}.  
\end{equation}
This means that we cannot use the standard direct methods like the Hirota 
approach and have to build solutions almost `from scratch'.

%%%%%%%%%%%%%%%%%%%%%%%%%%%%%%%%%%%%%%%%%%%%%%%%%%%%%%%%%%%%%%%%%%%%%%%%%%%%%%%
\section{Soliton matrices. \label{sec:matrices} }
%%%%%%%%%%%%%%%%%%%%%%%%%%%%%%%%%%%%%%%%%%%%%%%%%%%%%%%%%%%%%%%%%%%%%%%%%%%%%%%

In this section we construct solutions for the system \eref{syst:tosolve} 
from the soliton matrices %similar to the ones 
studied in \cite{V15}. 
Partly, the calculations presented here are similar to ones of \cite{V15}. 
However, this time we need more deep analysis of the properties of the 
soliton matrices: the results of \cite{V15} are not enough to tackle 
the quadrilinear restrictions discussed in the previous section.

%%%%%%%%%%%%%%%%%%%%%%%%%%%%%%%%%%%%%%%%%%%%%%%%%%%%%%%%%%%%%%%%%%%%%%%%%%%%%%%
\subsection{ Definitions. }
%%%%%%%%%%%%%%%%%%%%%%%%%%%%%%%%%%%%%%%%%%%%%%%%%%%%%%%%%%%%%%%%%%%%%%%%%%%%%%%

We define the soliton matrices by the so-called `rank one condition' 
\begin{equation}
  \begin{array}{l}
  \mathsf{\L2} \mathsf{\A1} -  \mathsf{\A1} \mathsf{\L1} 
  = 
  \keta1\braa1 
  \\[2mm]
  \mathsf{\L1} \mathsf{\A2} -  \mathsf{\A2} \mathsf{\L2} 
  = 
  \keta2\braa2 
  \end{array}
\end{equation}
where $\mathsf{\L1}$ and $\mathsf{\L2}$ are constant $N \times N$ diagonal 
matrices, 
$\keta1$ and $\keta2$ are constant $N$-columns 
while  
$\braa1$ and $\braa2$ are $N$-component rows that depend on the 
coordinates describing the model. 

For our purposes it is helpful to rewrite this equation as an intertwining 
relation
\begin{equation}
  \begin{array}{l}
  \left( \; \mathsf{\L2} - \keta1\brab1 \; \right) \mathsf{\A1} 
  = 
  \mathsf{\A1} \mathsf{\L1} 
  \\[2mm]
  \left( \; \mathsf{\L1} - \keta2\brab2 \right) \mathsf{\A2} 
  = 
  \mathsf{\A2} \mathsf{\L2} 
  \end{array}
\end{equation}
with \emph{constant} $N$-rows $\brab{1,2}$ which are defined as 
\begin{equation}
  \braa{i} = \brab{i} \mathsf{\A{0}_{i}}, \qquad (i=1,2). 
\end{equation}

The shifts $\Shift{}$ are \emph{defined} as the right multiplication 
\begin{equation}
  \begin{array}{l}
  \Shifted{\zeta}{ \mathsf{\A1}} 
  = 
  \mathsf{\A1} 
  \left( \mathsf{\L1} + \zeta \right) 
  \left( \mathsf{\L1} - \zeta \right)^{-1} 
  \\[2mm]
  \Shifted{\zeta}{ \mathsf{\A2}} 
  = 
  \mathsf{\A2} 
  \left( \mathsf{\L2} - \zeta \right) 
  \left( \mathsf{\L2} + \zeta \right)^{-1} 
  \end{array}
\label{evol:A}
\end{equation}
(we do not indicate the unit matrix explicitly and write 
$\mathsf{\L0} \pm \zeta$ instead of $\mathsf{\L0} \pm \zeta\mathsf{1}$, etc). 

%%%%%%%%%%%%%%%%%%%%%%%%%%%%%%%%%%%%%%%%%%%%%%%%%%%%%%%%%%%%%%%%%%%%%%%%%%%%%%%
\subsection{ One-shift formulae. }
%%%%%%%%%%%%%%%%%%%%%%%%%%%%%%%%%%%%%%%%%%%%%%%%%%%%%%%%%%%%%%%%%%%%%%%%%%%%%%%

From \eref{evol:A} one can derive the action of the shifts 
$\Shift{}$ on the determinants $\tau$ 
\begin{equation}
  \tau 
  = 
  \det\left| 1 + \mathsf{\A1}\mathsf{\A2} \right| 
\label{def:tau}
\end{equation}
and the inverse matrices 
\begin{equation}
  \begin{array}{l}
  \mathsf{\G1} 
  = 
  \left( 1 + \mathsf{\A1}\mathsf{\A2} \right)^{-1} 
  \\[2mm]
  \mathsf{\G2} 
  = 
  \left( 1 + \mathsf{\A2}\mathsf{\A1} \right)^{-1}. 
  \end{array}
\end{equation}
The corresponding formulae can be written as 
\begin{eqnarray}
  \frac{ \Shifted{\zeta}{\tau} }{ \tau } 
  & = & 
  1
  + 2 \zeta \u{1\zeta} \, 
    \brab{1\zeta} 
    \mathsf{\A1} \mathsf{\G2} \mathsf{\A2} 
    \keta{1\zeta} 
\label{evol:tau}
\\
  & = & 
  1 
  - 2 \zeta \u{2\zeta} \, 
    \brab{2\zeta} 
    \mathsf{\A2}\mathsf{\G1}\mathsf{\A1} 
    \keta{2\zeta} 
\end{eqnarray} 
and 
\begin{eqnarray}
  \frac{ \Shifted{\zeta}{\tau} }{ \tau } 
  \left( \Shift{\zeta} - 1 \right) \mathsf{\G1} 
  & = & 
  2 \zeta \u{2\zeta} \, 
    \mathsf{\G1}\mathsf{\A1} 
    \keta{2\zeta}\brab{2\zeta} 
    \mathsf{\G2}\mathsf{\A2} 
\\
  \frac{ \Shifted{\zeta}{\tau} }{ \tau } 
  \left( \Shift{\zeta} - 1 \right) \mathsf{\G2} 
  & = & 
  - 2 \zeta \u{1\zeta} \, 
    \mathsf{\G2}\mathsf{\A2} 
    \keta{1\zeta}\brab{1\zeta} 
    \mathsf{\G1}\mathsf{\A1} 
\label{evol:G}
\end{eqnarray}
where constants $\u{i\zeta}$ are given by 
\begin{equation}
  \u{i\zeta} 
  = 
  \frac{ 1 }{ 1 - \brab{i\,\zeta} \keta{i} }, 
  \qquad (i=1,2) 
\end{equation}
and 
\begin{equation}
  \begin{array}{l} 
  \brab{1\zeta} = \brab{1} \left( \mathsf{\L2} - \zeta \right)^{-1} 
  \\[2mm]
  \brab{2\zeta} = \brab{2} \left( \mathsf{\L1} + \zeta \right)^{-1} 
  \end{array}
\qquad
  \begin{array}{l}
  \keta{1\zeta} 
  = 
  \left( \mathsf{\L2} + \zeta \right)^{-1} \keta{1}
  \\[2mm]
  \keta{2\zeta} 
  = 
  \left( \mathsf{\L1} - \zeta\right)^{-1} \keta{2}. 
  \end{array}
\end{equation}

Introducing the new functions 
\begin{equation}
  \begin{array}{l}
  \q0 = 1 - \brab{1} \mathsf{\G1} \keta{1}, 
  \\[2mm]
  \q3 = 1 - \brab{2} \mathsf{\G2} \keta{2}, 
  \end{array}
\qquad
  \begin{array}{l}
  \q1 = \brab{1} \mathsf{\G1} \mathsf{\A1} \keta{2}, 
  \\[2mm]
  \q2 = \brab{2} \mathsf{\G2} \mathsf{\A2} \keta{1}, 
  \end{array}
\label{def:pqrs}
\end{equation}
one can derive from \eref{evol:A} and \eref{evol:G} 
\begin{equation}
  \frac{ \Shifted{\zeta}{\tau} }{ \tau } 
  \left( \Shift{\zeta} - 1 \right) 
  \left( \begin{array}{c} \q0 \\ \q1 \\ \q2 \\ \q3 \end{array} \right) 
  = 
  2 \zeta \u{\zeta} \; 
    \left( \begin{array}{r} 
      - \q1_{\zeta}\q2_{\zeta} \\ 
        \q1_{\zeta}\q3_{\zeta} \\ 
      - \q0_{\zeta}\q2_{\zeta} \\ 
        \q1_{\zeta}\q2_{\zeta} 
    \end{array} \right) 
\label{x:pqrs}
\end{equation}
and
\begin{equation}
  \frac{ \Shifted{\zeta}{\tau} }{ \tau } 
  = 
  \u{\zeta} \left( \q0_{\zeta}\q3_{\zeta} + \q1_{\zeta}\q2_{\zeta} \right) 
\label{x:tau}
\end{equation}
where 
\begin{equation}
  \u{\zeta} = \u{1\zeta}\u{2\zeta} 
\end{equation}
and 
\begin{equation}
  \begin{array}{l}
  \q0_{\zeta} = 1 - \brab{1\zeta} \mathsf{\G1} \keta{1} 
  \\[2mm]
  \q3_{\zeta} = 1 - \brab{2\zeta} \mathsf{\G2} \keta{2} 
  \end{array}
\qquad
  \begin{array}{l}
  \q1_{\zeta} = \brab{1\zeta} \mathsf{\G1} \mathsf{\A1} \keta{2} 
  \\[2mm]
  \q2_{\zeta} = \brab{2\zeta} \mathsf{\G2} \mathsf{\A2} \keta{1}. 
  \end{array}
\end{equation}

%%%%%%%%%%%%%%%%%%%%%%%%%%%%%%%%%%%%%%%%%%%%%%%%%%%%%%%%%%%%%%%%%%%%%%%%%%%%%%%
\subsection{ Two-shift formulae. }
%%%%%%%%%%%%%%%%%%%%%%%%%%%%%%%%%%%%%%%%%%%%%%%%%%%%%%%%%%%%%%%%%%%%%%%%%%%%%%%

By means of straightforward (although rather cumbersome) calculations based 
on \eref{evol:A} and \eref{evol:tau}--\eref{evol:G} one can describe the 
`evolution' of the functions $\q0_{\zeta}, ..., \q3_{\xi}$, 
\begin{equation}
  \frac{ \Shifted{\eta}{\tau} }{ \tau } 
  \left( \Shifted{\eta} - 1 \right) 
  \left( \begin{array}{l}
    \q0_{\xi} \\ \q1_{\xi} \\ \q2_{\xi} \\ \q3_{\xi} 
  \end{array} \right) 
  = 
  \frac{ 2 \eta \u{\eta} }{ \xi - \eta }
  \left( \begin{array}{l}
    \q0_{\xi} \q1_{\eta} \q2_{\eta} 
    - 
    \q1_{\xi} \q0_{\eta} \q2_{\eta} 
    \\
    \q1_{\xi} \q0_{\eta} \q3_{\eta} 
    - 
    \q0_{\xi} \q1_{\eta} \q3_{\eta} 
    \\
    \q2_{\xi} \q0_{\eta} \q3_{\eta} 
    - 
    \q3_{\xi} \q0_{\eta} \q2_{\eta} 
    \\
    \q3_{\xi} \q1_{\eta} \q2_{\eta} 
    - 
    \q2_{\xi} \q1_{\eta} \q3_{\eta} 
  \end{array} \right), 
\label{x:pqrsz}
\end{equation}
and to obtain the following two-shift identity for the tau-functions:
\begin{equation}
  \tau \Shifted[1]{\xi\eta}{ \tau} 
  - 
  \Shifted[1]{\xi}{\tau} \Shifted[1]{\eta}{\tau} 
  = 
  \frac{ 4 \xi \eta \u{\xi} \u{\eta} }{ (\xi - \eta)^{2} } 
  \left( \q0_{\xi} \q1_{\eta} -  \q0_{\eta} \q1_{\xi} \right)
  \left( \q2_{\eta} \q3_{\xi} -  \q2_{\xi} \q3_{\eta} \right) 
  \tau^{2}. 
\label{xx:tau}
\end{equation}

Equations \eref{x:pqrs} together with \eref{x:pqrsz}  lead to 
\begin{equation}
  \begin{array}{lcl}
  \Shifted{\xi\eta}{ \q0} - \q0 + \xi + \eta 
  & = & 
  \frac{\xi+\eta}{\xi-\eta} \; f_{\xi\eta} \left(
    \Shifted{\xi}{\q0} - \Shifted{\eta}{\q0} + \xi - \eta
  \right) 
  \\[2mm]
  \Shifted{\xi\eta}{\q1} - \q1 
  & = & 
  \frac{\xi+\eta}{\xi-\eta} \; f_{\xi\eta} 
  \left( \Shifted{\xi}{\q1} - \Shifted{\eta}{\q1} \right) 
  \\[2mm]
  \Shifted{\xi\eta}{\q2} - \q2 
  & = & 
  \frac{\xi+\eta}{\xi-\eta} \; f_{\xi\eta} 
  \left( \Shifted{\xi}{\q2} - \Shifted{\eta}{\q2} \right) 
  \end{array}
\label{xx:pqr}
\end{equation}
where 
\begin{equation}
  f_{\xi\eta} 
  = 
  \frac{ \Shifted[1]{\xi}{\tau} \Shifted[1]{\eta}{\tau} }
       { \tau \Shifted[1]{\xi\eta}{\tau} }. 
\end{equation}
We do not write similar expression for $\q3$ because, as follows from 
\eref{x:pqrs}, 
$\Shift{\zeta} \left( \q0 + \q3 \right) =  \q0 +  \q3$, 
which means that 
$\q0 + \q3 =  constant$. 

Introducing the new function 
\begin{equation}
  \q4 = \q0 + \chi 
\label{def:w}
\end{equation}
where $\chi$ is the `linear' function defined by 
\begin{equation}
  \Shifted{\zeta}{\chi} = \chi + \zeta 
\label{def:chi}
\end{equation}
one can rewrite \eref{xx:pqr} as 
\begin{equation}
  \left( \Shift{\xi\eta} - 1 \right) 
  \left( \begin{array}{c} \q1 \\ \q2 \\ \q4 \end{array} \right)
  = 
  \case{\xi + \eta}{\xi - \eta} \; 
  f_{\xi\eta} 
  \left( \Shift{\xi} - \Shift{\eta} \right) 
  \left( \begin{array}{c} \q1 \\ \q2 \\ \q4 \end{array} \right).
\label{sh:qrw}
\end{equation}

Finally, these equations together with \eref{xx:tau}, \eref{x:pqrs} and 
\eref{x:tau} yield 
\begin{equation}
  \left( \Shifted{\xi\eta}{\q4} -  \q4 \right)^{2} 
  - 
  \left( \Shifted{\xi\eta}{\q1} - \q1 \right) 
  \left( \Shifted{\xi\eta}{\q2} - \q2 \right) 
  = 
  (\xi + \eta)^{2} f_{\xi\eta}. 
\label{norm:qrw}
\end{equation}

It is easy to note that the last two equations have the structure of 
system \eref{syst:tosolve} with 
$A_{\xi\eta} = (\xi - \eta)/(\xi + \eta)$, 
$B_{\xi\eta} = 1/(\xi + \eta)^{2}$ 
and hence
$\varepsilon_{\xi\eta} = \xi^{2} - \eta^{2}$.
The only difference is that the quadratic form in \eref{norm:qrw} is not the 
Euclidean norm of the vector $(q,r,w)^{T}$.
Thus, the last problem we have to solve is to construct, 
of the functions $\q1$, $\q2$ and $\q4$, the vectors $\vec{\phi}$ with the appropriate
norm.

%%%%%%%%%%%%%%%%%%%%%%%%%%%%%%%%%%%%%%%%%%%%%%%%%%%%%%%%%%%%%%%%%%%%%%%%%%%%%%%
\subsection{ Involution. }
%%%%%%%%%%%%%%%%%%%%%%%%%%%%%%%%%%%%%%%%%%%%%%%%%%%%%%%%%%%%%%%%%%%%%%%%%%%%%%%

Till now, we have not specified whether the functions introduced in this 
section are real or complex. All formulae presented above are suitable for 
both cases. Here, we discuss the symmetry of the soliton matrices with respect 
to the comlex conjugation.

It is easy to verify that the restrictions 
\begin{equation}
  \mathsf{\L2} = \overline{\mathsf{\L1}}, \quad
  \brab{2} = \overline{\brab{1}}, \quad
  \keta{2} = \overline{\keta{1}}, 
\label{inv:L}
\end{equation}
where the overbar stands for the complex conjugation, lead to 
\begin{equation}
  \mathsf{\A2} = \overline{\mathsf{\A1}}. 
\label{inv:A}
\end{equation}
It follows from \eref{evol:A} that to ensure the consistency of the action of 
the shifts $\Shift{\zeta}$ with the involution \eref{inv:A} we have to restrict 
ourselves with pure imaginary $\zeta$, 
\begin{equation}
  \Re\zeta = 0 
  \quad\Rightarrow\quad
  \Shifted{\zeta}{\mathsf{\A2}} 
  = 
  \overline{\Shifted{\zeta}{\mathsf{\A1}}}. 
\end{equation}
Hereafter, we use the `real' shifts $\ShiftR{}$ defined by 
\begin{equation}
	\ShiftR\lambda = \Shift{i\lambda}, \qquad (\Im\lambda=0). 
\end{equation}
One can derive from \eref{inv:L}, \eref{inv:A} and the definitions 
\eref{def:pqrs} the identities
\begin{equation}
  \q3 = \overline{\q0}, \quad \q2 = \overline{\q1} 
\label{inv:pqrs}
\end{equation}
which are compatible with the action of the shifts $\ShiftR{\lambda}$, 
\begin{equation}
  \ShiftR\lambda\q3 = \overline{\ShiftR\lambda\q0}, \quad 
  \ShiftR\lambda\q2 = \overline{\ShiftR\lambda\q1}. 
\end{equation}
We have already mentioned that $\q0 + \q3$ is constant with respect to the 
shifts. In the context of \eref{inv:pqrs}, this reads 
\begin{equation}
  \left( \ShiftR\lambda - 1 \right)\q0 
  = 
  i \left( \ShiftR\lambda - 1 \right)\Im\q0 
\end{equation}
which, together with the definition \eref{def:chi}, implies
\begin{equation}
  \left( \ShiftR\lambda - 1 \right)\q4 
  = 
  i \left( \ShiftR\lambda - 1 \right)\Im\q4. 
\end{equation}
Now, we can rewrite equation \eref{norm:qrw} in terms of $\q1$ and $\q4$ 
\begin{equation}
  \left( \ShiftR{\lambda\mu}\Im\q4 -  \Im\q4 \right)^{2} 
  + 
  \left| \ShiftR{\lambda\mu}\q1 - \q1 \right|^{2}
  = 
  (\lambda + \mu)^{2} f^{\scriptscriptstyle{R}}_{\lambda\mu} 
\end{equation}
where 
$f^{\scriptscriptstyle{R}}_{\lambda\mu} 
  = 
  \left( \ShiftR{\lambda}\tau \right) \left( \ShiftR{\mu}\tau \right) / 
  \tau \left( \ShiftR{\lambda\mu}\tau \right) 
$.

Thus, we can formulate the main result of this section.

\begin{proposition}
\label{prop:main}
Vector $\vec{\phi}$ defined as 
\begin{equation}
  \vec{\phi} 
  = 
  \left(\Re\q1, \; \Im\q1, \; \Im\q4 \right)^{T} 
\end{equation}
with functions $\q1$, $\q2$ and $\q4$ defined in 
\eref{def:pqrs}, \eref{def:w} and \eref{def:chi} satisfies 
\begin{equation}
  \ShiftR\mu\vec{\phi} - \ShiftR\nu\vec{\phi} 
  = 
  \left( \mu^{2} - \nu^{2} \right) 
  \frac{ \ShiftR{\mu\nu}\vec{\phi} - \vec{\phi} }
       { \left\| \ShiftR{\mu\nu}\vec{\phi} - \vec{\phi} \right\|^{2} }
\end{equation}
with arbitrary real $\mu$ and $\nu$.
\end{proposition}

%%%%%%%%%%%%%%%%%%%%%%%%%%%%%%%%%%%%%%%%%%%%%%%%%%%%%%%%%%%%%%%%%%%%%%%%%%%%%%%%
\section{N-soliton solutions. \label{sec:solitons} }
%%%%%%%%%%%%%%%%%%%%%%%%%%%%%%%%%%%%%%%%%%%%%%%%%%%%%%%%%%%%%%%%%%%%%%%%%%%%%%%%

%%%%%%%%%%%%%%%%%%%%%%%%%%%%%%%%%%%%%%%%%%%%%%%%%%%%%%%%%%%%%%%%%%%%%%%%%%%%%%%%
\subsection{Vector discrete KdV equation. }
%%%%%%%%%%%%%%%%%%%%%%%%%%%%%%%%%%%%%%%%%%%%%%%%%%%%%%%%%%%%%%%%%%%%%%%%%%%%%%%%

As follows from proposition \ref{prop:main}, to obtain soliton solutions for 
\eref{eq:vkdv} we have to make two simple steps. First, we introduce the 
dependence on $m$ and $n$ as 
\begin{equation}
  \vec{\phi}_{m,n} 
  = 
  \left(\ShiftR\mu\right)^{m}
  \left(\ShiftR\nu\right)^{n}
  \vec{\phi}.
\end{equation}
Secondly, we have to rescale $\vec{\phi}_{m,n}$ in order to make the 
right-hand side of \eref{eq:vkdv} equal to unity,
\begin{equation}
  \vec{\phi}_{m,n} 
  \to 
  \left| \mu^{2} - \nu^{2} \right|^{-1/2} \vec{\phi}_{m,n}. 
\end{equation}
After that, we can present the $N$-soliton solutions for 
%the vector discrete KdV equation 
\eref{eq:vkdv} as follows.

%%%%%%%%%%%%%%%%%%%%%%%%%%%%%%%%%%%%%%%%%%%%%%%%%%%%%%%%%%%%%%%%%%%%%%%%%%%%%%%%
\begin{proposition} \label{prop:vkdv}
The $N$-soliton solutions for the vector discrete KdV equation \eref{eq:vkdv} 
can be presented as 
\begin{equation}
  \vec{\phi}_{m,n} 
  = 
  \vec{\phi}^{\rm bg}_{m,n} + \vec{\phi}^{\rm sol}_{m,n} 
\end{equation}
where the background part, $\vec{\phi}^{\rm bg}_{m,n}$ 
is the linear function of $m$ and $n$, 
\begin{equation}
  \vec{\phi}^{\rm bg}_{m,n} 
  = 
  \frac{ m\mu + n\nu }{ \left| \mu^{2} - \nu^{2} \right|^{1/2} } 
  {\scriptstyle \left(\begin{array}{c} 0 \\[-1mm] 0 \\[-1mm] 1   \end{array}\right) } 
\end{equation} 
and 
\begin{equation}
  \vec{\phi}^{\rm sol}_{m,n} 
  = 
  \frac{ 1 }{ \left| \mu^{2} - \nu^{2} \right|^{1/2} } 
  \left(\begin{array}{c}
  \Re\brab{} \mathsf{G}_{m,n} \mathsf{A}_{m,n} \keta{}   %2  
  \\[1mm] 
  \Im\brab{} \mathsf{G}_{m,n} \mathsf{A}_{m,n} \keta{}   %2  
  \\[1mm] 
  - \Im \brab{} \mathsf{G}_{m,n} \keta{}  
  \end{array}\right). 
\label{sls:vkdv}
\end{equation}
Here 
\begin{equation}
  \mathsf{A}_{m,n} 
  = 
  \mathsf{A} \; 
  \mathsf{H}_{\mu}^{m} 
  \mathsf{H}_{\nu}^{n} 
\end{equation}
with the constant matrices $\mathsf{A}$ and $\mathsf{H}_{\mu,\nu}$ given by 
\begin{equation}
  \mathsf{A} 
  = 
  \left( 
     \frac{ a_{k} }{ \bar{L}_{j} - L_{k} } 
  \right)_{j,k= 1,...,N} 
\end{equation}
\begin{equation}
  \mathsf{H}_{\lambda} 
  = 
  \mbox{\rm diag} \left( 
    \frac{ L_{k} + i\lambda }{ L_{k} - i\lambda } 
  \right)_{k= 1,...,N}, 
\label{def:H}
\end{equation}
and 
\begin{equation}
  \mathsf{G}_{m,n} 
  = 
  \left( 1 + \mathsf{A}_{m,n}\overline{\mathsf{A}_{m,n}} \; \right)^{-1}. 
\end{equation}
The constant $N$-row $\brab{}$ is defined by 
$ \brab{} 
  = 
  \left( \beta_{1}, ..., \beta_{N} \right) 
  = 
  \left( a_{1}, ..., a_{N} \right)\mathsf{A}^{-1} 
$, 
the $N$-column $\keta{}$ is defined as 
$\keta{} = \left( 1, ..., 1 \right)$ 
and $\{ a_{k},L_{k} \}_{k=1,\dots,N}$
and $\mu,\nu$ are arbitrary constants. 

\end{proposition}
%%%%%%%%%%%%%%%%%%%%%%%%%%%%%%%%%%%%%%%%%%%%%%%%%%%%%%%%%%%%%%%%%%%%%%%%%%%%%%%%

Note that we use $L_{j}$ for the elements of the diagonal matrix $\mathsf{L}$, 
\begin{equation}
  \mathsf{L} = \mbox{diag}\left( L_{1}, \dots, L_{N} \right) 
\end{equation}
and that we have eliminated some `redundant' constants by replacing 
$\keta{1,2}$ with $\keta{}$ 
(the components of the columns $\keta{1,2}$ can be `included' in the arbitrary 
constants $a_{k}$).

In the one soliton case ($N=1$) the matrix $\mathsf{L}$ becomes a scalar, 
$\mathsf{L}\to L$ and we have only one $a$-parameter, $a=a_{1}$. 
The formulae from proposition \ref{prop:vkdv} can be rewritten as
\begin{equation}
  \vec{\phi}^{\rm sol}_{m,n} 
  = 
  \frac{ \rho }{ \cosh h_{m,n} } 
  \left( 
  \cos \varphi_{m,n}, \; 
  \sin \varphi_{m,n}, \; 
  e^{-h_{m,n}} 
  \right)^{T} 
\label{soliton:vkdv}
\end{equation}
where 
$\rho = |\Im L| / \left| \mu^{2} - \nu^{2} \right|^{1/2}$ 
and 
$h_{m,n}$ and $\varphi_{m,n}$ are linear functions of $m$ and $n$,   

\begin{eqnarray}
  h_{m,n} 
  & = & 
    \kappa_{R}(\mu) \, m
  + \kappa_{R}(\mu) \, n
  + h_{*}
\\
  \varphi_{m,n} 
  & = & 
    \kappa_{I}(\mu) \, m
  + \kappa_{I}(\nu) \, n 
  + \varphi_{*}
\end{eqnarray}
where 
\begin{equation}
  \kappa_{R}(\lambda) = \ln\left|\frac{L + i\lambda}{L - i\lambda}\right|, 
  \quad
  \kappa_{I}(\lambda) = \arg\frac{L + i\lambda}{L - i\lambda},
\label{def:kappa}
\end{equation}
\begin{equation}
  h_{*} = \ln\left|\frac{a}{2\Im L}\right|,  
  \quad
  \varphi_{*} = \arg a.
\label{def:cstar}
\end{equation}
Calculating the norm of $\vec{\phi}^{\rm sol}_{m,n}$, 
\begin{equation}
  \left\| \vec{\phi}^{\rm sol}_{m,n} \right\|^{2} 
  = 
  \frac{ 4\rho^{2} }{ 1 + e^{2h_{m,n}} } 
  \quad \to \quad
  \left\{\begin{array}{rcl}
  4 \rho^{2} & \mbox{as} & h_{m,n} \to - \infty
  \\
  0 & \mbox{as} & h_{m,n} \to + \infty
  \end{array}\right. 
\end{equation}
one can see that the obtained line soliton has a step- or kink-like structure: 
the $\left\| \vec{\phi}^{\rm sol}_{m,n} \right\|$ is bounded between $0$ 
(which it attains in one asymptotic direction) and $2\rho$
(which it attains in the opposite direction).
However, the part of $\vec{\phi}$ which is perpendicular to 
$\vec{\phi}^{\rm bg}$ (the first two components in \eref{soliton:vkdv}) 
reveals typical soliton $sech$-behaviour.

\begin{figure}%
\begin{center}
\includegraphics{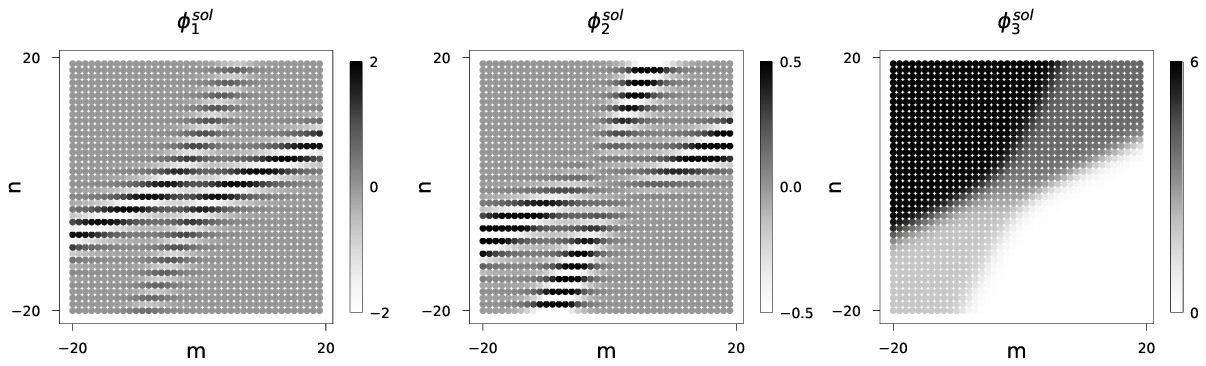}%
\end{center}
\caption{%
The $(m,n)$-dependence of the components of the two-soliton solution 
\eref{sls:vkdv}.
}%
\label{fig:1}%
\end{figure}
To illustrate the structure of the two-soliton solutions we calculate 
\eref{sls:vkdv} for some fixed set of soliton parameters:
$L_{1} = 0.1 + i$, $L_{2} = 0.1 + 2i$, $a_{1} = 10$, $a_{2} = 9$.
To make the plots more clear we present in figure \ref{fig:1} only the soliton 
part of the solution, $\vec{\phi}^{\rm sol}$. 
As in the one-soliton case, we can see that the third component of $\vec{\phi}$ 
(the part of $\vec{\phi}$ which is parallel to $\vec{\phi}^{\rm bg}$) 
has the two-kink structure, while the first two 
(the part of $\vec{\phi}$ which is perpendicular to $\vec{\phi}^{\rm bg}$) 
have the stucture of two solitons (with sign-alternation along one of the 
directions).

%%%%%%%%%%%%%%%%%%%%%%%%%%%%%%%%%%%%%%%%%%%%%%%%%%%%%%%%%%%%%%%%%%%%%%%%%%%%%%%%
\subsection{Vector Yamilov equation. }
%%%%%%%%%%%%%%%%%%%%%%%%%%%%%%%%%%%%%%%%%%%%%%%%%%%%%%%%%%%%%%%%%%%%%%%%%%%%%%%%

To obtain the solitons of equation \eref{eq:vnve} 
using the result of proposition \ref{prop:main} 
we have to introduce the 
continuous variable $t$ so that the differentiating $d/dt$ reproduces the 
action of the operator \eref{def:D} or 
$
  \frac{ 1 }{\mu^{2} - \nu^{2}} 
  \left( \ShiftR{\mu} \left(\ShiftR{\nu}\right) ^{-1} - 1 \right)  
$. 
One can obtain from \eref{evol:A} that 
\begin{equation}
  \left( \ShiftR{\mu} \left(\ShiftR{\nu}\right) ^{-1} - 1 \right) \mathsf{A} 
  = 
  2i (\mu - \nu) \; 
  \mathsf{A} \, 
  \mathsf{L}
  \left( \mathsf{L} - i\mu \right)^{-1} 
  \left( \mathsf{L} + i\nu \right)^{-1} 
\end{equation}
which leads to 
\begin{equation}
  \frac{ d }{ d t } \, \mathsf{A}(t) 
  = 
  i 
  \mathsf{A}(t) \; 
  \mathsf{L}
  \left( \mathsf{L}^{2} + \nu^{2} \right)^{-1} 
\end{equation}
or
\begin{equation}
  \mathsf{A}(t) = 
  \mathsf{A}(0) \; 
  \exp\left( i \mathsf{\Omega} t \right),
  \qquad 
  \mathsf{\Omega} 
  = 
  \mathsf{L}
  \left( \mathsf{L}^{2} + \nu^{2} \right)^{-1} 
\end{equation}
The $n$-dependence of the matrices $\mathsf{A}$ 
(and, hence, of $\vec{\phi}$) 
is governed, as in the previous section, by the matrix 
$\mathsf{H} = \mathsf{H}_{\nu}$ from \eref{def:H}. 
Thus, we have all necessary to present the solitons of \eref{eq:vnve}.

%%%%%%%%%%%%%%%%%%%%%%%%%%%%%%%%%%%%%%%%%%%%%%%%%%%%%%%%%%%%%%%%%%%%%%%%%%%%%%%%
\begin{proposition}
The $N$-soliton solutions for the vector Yamilov equation \eref{eq:vnve} 
can be presented as 
\begin{equation}
  \vec{\phi}_{n}(t) 
  = 
  \vec{\phi}^{\rm bg}_{n}(t) + \vec{\phi}^{\rm sol}_{n}(t) 
\end{equation}
where the background part, $\vec{\phi}^{\rm bg}_{n}(t)$ 
is the linear function of $t$ and $n$, 
\begin{equation}
  \vec{\phi}^{\rm bg}_{m,n} 
  = 
  \left( \frac{t}{2\nu} +n\nu \right) 
  {\scriptstyle \left(\begin{array}{c} 0 \\[-1mm] 0 \\[-1mm] 1   \end{array}\right) } 
\end{equation} 
and
\begin{equation}
  \vec{\phi}^{\rm sol}_{n}(t) 
  = 
  \left(\begin{array}{c}
  \Re\brab{} \mathsf{G}_{n}(t) \mathsf{A}_{n}(t) \keta{} 
  \\[2mm] 
  \Im\brab{} \mathsf{G}_{n}(t) \mathsf{A}_{n}(t) \keta{} 
  \\[1mm] 
  - \Im \brab{} \mathsf{G}_{n}(t) \keta{} %+ \frac{t}{2\nu} +n\nu
  \end{array}\right). 
\label{sls:vnve}
\end{equation}
Here 
\begin{equation}
  \mathsf{A}_{n}(t) 
  = 
  \mathsf{A} \; 
  \mathsf{H}^{n} 
  \exp\left( i \mathsf{\Omega} t \right),
\end{equation}
with the constant matrices $\mathsf{A}$, $\mathsf{H}$ and $\mathsf{\Omega}$ 
given by 
\begin{equation}
  \mathsf{A} 
  = 
  \left(\; 
     \frac{ a_{k} }{ \bar{L}_{j} - L_{k} } 
  \;\right)_{j,k= 1,...,N}
\end{equation}
\begin{equation}
  \mathsf{H} 
  = 
  \mbox{\rm diag} \left(\; 
    \frac{ L_{k} + i\nu }{ L_{k} - i\nu } 
  \;\right)_{k= 1,...,N}, 
\end{equation}
\begin{equation}
  \mathsf{\Omega} 
  = 
  \mbox{\rm diag} \left( \; 
    \frac{ L_{k} }{ L_{k}^{2} + \nu^{2} } 
  \;\right)_{k= 1,...,N}, 
\end{equation}
and 
\begin{equation}
  \mathsf{G}_{n}(t) 
  = 
  \left( 1 + \mathsf{A}_{n}(t)\overline{\mathsf{A}_{n}(t)} \right)^{-1}. 
\end{equation}
The constant $N$-row $\brab{}$ is defined by 
$ \brab{} 
  = 
  \left( \beta_{1}, ..., \beta_{N} \right) 
  = 
  \left( a_{1}, ..., a_{N} \right)\mathsf{A}^{-1} 
$, 
the $N$-column $\keta{}$ is defined as 
$\keta{} = \left( 1, ..., 1 \right)$ 
and $\{ a_{k},L_{k} \}_{k=1,\dots,N}$
and $\nu$ are arbitrary constants. 
\end{proposition}
%%%%%%%%%%%%%%%%%%%%%%%%%%%%%%%%%%%%%%%%%%%%%%%%%%%%%%%%%%%%%%%%%%%%%%%%%%%%%%%%

Clearly, the structure of the one soliton solution is the same as in the case 
of the vector discrete KdV equation, 
\begin{equation}
  \vec{\phi}^{\rm sol}_{n}(t) 
  = 
  \frac{ \rho }{ \cosh h_{n}(t) } 
  \left( 
  \cos \varphi_{n}(t), \; 
  \sin \varphi_{n}(t), \; 
  e^{-h_{n}(t)} 
  \right)^{T}. 
\end{equation}
The differences are in that 
$\rho = |\Im L|$ 
and in the `dispersion laws',
\begin{eqnarray}
  h_{n}(t) 
  & = & 
  - \gamma t 
  + \kappa_{R}(\nu) n 
  + h_{*}, 
\\
  \varphi_{n}(t) 
  & = & 
  \omega t  
  + \kappa_{I}(\nu) n 
  + \varphi_{*}
\end{eqnarray}
where 
\begin{equation}
  \omega = - \frac{\Im a}{2\Im L}, 
  \qquad
  \gamma = \frac{\Re a}{2\Im L} 
\end{equation}
while the functions $\kappa_{R,I}(\nu)$ and the constants 
$h_{*}$ and $\varphi_{*}$ are defined in \eref{def:kappa} and \eref{def:cstar}. 

\begin{figure}%
\begin{center}
\includegraphics{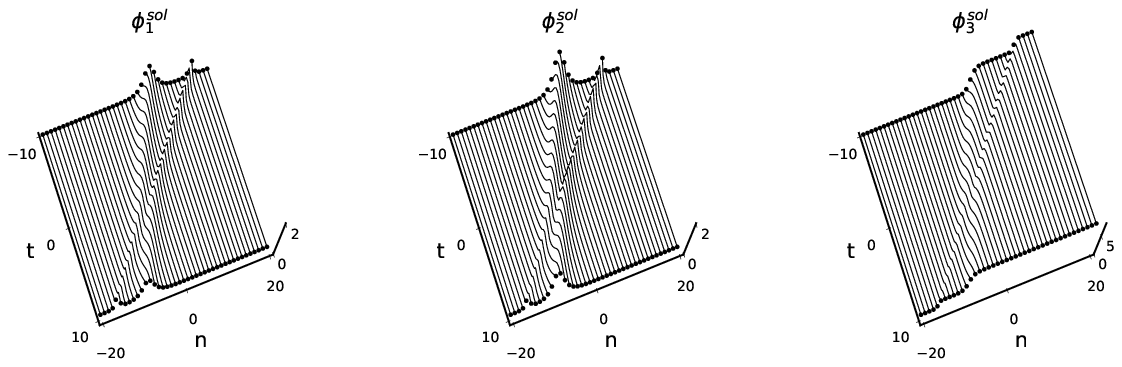}%
\end{center}
\caption{%
The $(t,n)$-dependence of the components of the two-soliton solution \eref{sls:vnve}.
}%
\label{fig:2}%
\end{figure}

The two-soliton solution for 
$L_{1} = i$, $L_{2} = 2i$, $a_{1} = 2 + 2i$, $a_{2} = 2 + 3i$ 
and $\nu = -0.8$ is presented in figure \ref{fig:2}. 
Again, 
the part of $\vec{\phi}$ which is perpendicular to $\vec{\phi}^{\rm bg}$ has 
the structure of two $sech$--solitons, while the part of $\vec{\phi}$ which is 
parallel to $\vec{\phi}^{\rm bg}$ reveals the two--kink behaviour. 

%%%%%%%%%%%%%%%%%%%%%%%%%%%%%%%%%%%%%%%%%%%%%%%%%%%%%%%%%%%%%%%%%%%%%%%%%%%%%%%
\section{Discussion.}
%%%%%%%%%%%%%%%%%%%%%%%%%%%%%%%%%%%%%%%%%%%%%%%%%%%%%%%%%%%%%%%%%%%%%%%%%%%%%%%

To conclude, we would like to stress out once more the main difference between 
the calculations of this work and other our works devoted to solitons of the 
vector lattice models, for example, \cite{V16,V19}.
In \cite{V16,V19}, our starting point was some \emph{scalar} identitities for 
the soliton matrices from \cite{V15}. These identies were enough to 
(i) derive the vector ones, similar to equation \eref{eq:bimoutard}, 
or the first equation from \eref{syst:tosolve}, 
and (ii) to tackle the rectrictions similar to the second equation 
from \eref{syst:tosolve}. 
Here, the situation was more complicated: we had to return to the matrices 
discussed in \cite{V15} and to derive some aditional identities 
(absent in \cite{V15}), which are less `universal' but which gave us 
possibility to construct solitons for the models discussed in this paper.

Finally, according the so-called Hirota's three-soliton test
\cite{H87a,H87b,H87c,RGB89}, 
existence of $N$-soliton solutions can be viewed as an indication of the 
integrability of the models \eref{eq:vkdv} and \eref{eq:vnve}. 
Thus, a natural continuation of this work is to study the corresponding range 
problems mentioned in the Introduction (the Lax representation, 
conservation laws, Hamiltonian structures \textit{etc}). 
However these questions are out of the scope of this paper and may be 
considered in the following studies.

%%%%%%%%%%%%%%%%%%%%%%%%%%%%%%%%%%%%%%%%%%%%%%%%%%%%%%%%%%%%%%%%%%%%%%%%%%%%%%%
\section*{References} 

%\bibliography{vnve}{}

\begin{thebibliography}{10}
\expandafter\ifx\csname url\endcsname\relax
  \def\url#1{{\tt #1}}\fi
\expandafter\ifx\csname urlprefix\endcsname\relax\def\urlprefix{URL }\fi
\providecommand{\eprint}[2][]{\url{#2}}
% Bibliography created with iopart-num v2.1
% /biblio/bibtex/contrib/iopart-num

\bibitem{CWN86}
Capel H~W, Wiersma G~L and Nijhoff F~W 1986 {\em Physica A\/} {\bf 138} 76--99

\bibitem{PNC90}
Papageorgiou V~G, Nijhoff F~W and Capel H~W 1990 {\em Phys. Lett. A\/} {\bf
  147} 106--114

\bibitem{CNP91}
Capel H~W, Nijhoff F~W and Papageorgiou V~G 1991 {\em Phys. Lett. A\/} {\bf
  155} 377--387

\bibitem{NC95}
Nijhoff F and Capel H 1995 {\em Acta Appl. Math.\/} {\bf 39} 133--158

\bibitem{ABS03}
Adler V~E, Bobenko A~I and Suris Y~B 2003 {\em Commun. Math. Phys.\/} {\bf 233}
  513--543

\bibitem{Y83}
Yamilov R~I 1983 {\em Uspekhi Mat. Nauk\/} {\bf 38} 155--156

\bibitem{Y06}
Yamilov R 2006 {\em J. Phys. A\/} {\bf 39} R541--R623

\bibitem{MSY87}
Mikhailov A~V, Shabat A~B and Yamilov R~I 1987 {\em Russian Mathematical
  Surveys\/} {\bf 42} 1--63

\bibitem{PV03}
Pritula G~M and Vekslerchik V~E 2003 {\em J. Phys. A\/} {\bf 36} 213--226

\bibitem{LPSY08}
Levi D, Petrera M, Scimiterna C and Yamilov R 2008 {\em SIGMA\/} {\bf 4} 077

\bibitem{V15}
Vekslerchik V~E 2015 {\em J. Phys. A\/} {\bf 48} 445204

\bibitem{M82}
Miwa T 1982 {\em Proc. Japan Acad. ser.A\/} {\bf 58} 9--12

\bibitem{V16}
Vekslerchik V 2016 {\em J. Phys. A\/} {\bf 49} 455202

\bibitem{V19}
Vekslerchik V 2019 {\em SIGMA\/} {\bf 15} 028

\bibitem{H87a}
Hietarinta J 1987 {\em J. Math. Phys.\/} {\bf 28} 1732--1742

\bibitem{H87b}
Hietarinta J 1987 {\em J. Math. Phys.\/} {\bf 28} 2094--2101

\bibitem{H87c}
Hietarinta J 1987 {\em J. Math. Phys.\/} {\bf 28} 2586--2592

\bibitem{RGB89}
Ramani A, Grammaticos B and Bountis T 1989 {\em Physics Reports\/} {\bf 180}
  159--245

\end{thebibliography}
%\bibliographystyle{iopart-num}

\providecommand{\newblock}{}

%%%%%%%%%%%%%%%%%%%%%%%%%%%%%%%%%%%%%%%%%%%%%%%%%%%%%%%%%%%%%%%%%%%%%%%%%%%%%%%%
%%%%%%%%%%%%%%%%%%%%%%%%%%%%%%%%%%%%%%%%%%%%%%%%%%%%%%%%%%%%%%%%%%%%%%%%%%%%%%%%
\end{document}